\documentclass{elsart}
\usepackage{cite}
\usepackage{dcolumn}
\usepackage{graphicx}

\begin{document}

\begin{frontmatter}

\title{Wronskian method and the Schr\"{o}dinger eigenvalue march}
\author{Francisco M. Fern\'{a}ndez}
\address{INIFTA (UNLP, CCT La Plata-CONICET), Divisi\'{o}n Qu\'{i}mica Te\'{o}rica,\\
Blvd. 113 y 64 (S/N), Sucursal 4, Casilla de Correo 16,\\
1900 La Plata, Argentina}

\thanks[FMF]{e--mail: fernande@quimica.unlp.edu.ar}

\begin{abstract}
We compare the Wronskian method (WM) and the Schr\"odinger
eigenvalue march or canonical function method (SEM--CFM) for the
calculation of the energies and eigenfunctions of the
Schr\"odinger equation. The Wronskians between linearly
independent solutions of the Schr\"odinger equation provide a
rigorous basis for some of the assumptions of the SEM--CFM, like,
for example, the concept of ``saturation''. We compare the
performance of both approaches on a simple one--dimensional model
and suggest that taking into account the asymptotic behaviour of
the wavefunction (as is already done in the WM) may make the
SEM--CFM more efficient from a numerical point of view.
\end{abstract}

\end{frontmatter}

\section{Introduction}

\label{sec:intro}

In a recent paper Tannous and Langlois\cite{TL11} proposed the
Schr\"{o}dinger eigenvalue march (SEM) for the calculation of
eigenvalues of the Schr\"{o}dinger equation. The SEM was developed
by Kobeissi\cite{K82} under the name of canonical function method
(CFM) and even Tannous and Langlois\cite{TL10} and Tannous et
al\cite{TFL08} used to call it that way before renaming it.
According to the authors, the SEM compares favourably to the
well-known Numerov and shooting methods\cite {TL11}.
Leubner\cite{L83} argued that the underlying idea of the CFM is
very old and can be found in standard textbooks on differential
equations as the reduction of two--point boundary value problems
to initial value problems. In addition to it, he also proved the
equivalence between the CFM and the widely used shooting method.
In a rather late reply, Kobeissi\cite{K86} tried to show that the
CFM is considerably more accurate that the shooting method and
other algorithms. However, Tellinghuisen\cite{T88} confirmed
Leubner's conclusions and showed that the apparent numerical
advantage found by Kobeissi\cite{K86} was merely due to different
numerical precision in the results compared by this author.
Leubner's and Tellinghuisen's papers have been utterly omitted in
all later applications of the CFM\cite{TFL08}. It is worth quoting
Tellinghuisen's comment that `Although the CFM is fundamentally
less efficient than the Cooley algorithm, it does offer some
advantages in practical applications, as it avoids unnecessary
integration in the nonclassical region'.

We have recently discussed a method for the calculation of bound
states and transmission probabilities for one--dimensional wells
and barriers\cite {F11c,F11b,F11d}. It is based on the fact that
the coefficients of the linearly independent asymptotic
contributions to the wavefunction can be easily expressed in terms
of Wronskians. Those papers were mainly focused on the pedagogical
value of the Wronskian method (WM) that is already known since
long ago (see the references cited there \cite {F11c,F11b,F11d}).
Since the WM is also based on the so called canonical
functions\cite{K82,K86,TL10,TFL08} (or normalized
solutions\cite{TL11}) we think that it may be interesting and
fruitful from a pedagogical point of view to compare it with the
SEM (or CFM). In Sec.~\ref{sec:Wronskian} we outline the main
ideas behind the WM. In Sec.~\ref{sec:canonical} we discuss the
SEM--CFM and provide a rigorous basis for some of its equations by
means of the WM. In Sec.~\ref {sec:examples} we test, discuss and
verify the general results by means of simple examples. Finally,
in Sec.~\ref{sec:conclusions} we summarize the main results and
draw conclusions.

\section{The Wronskian method}

\label{sec:Wronskian}

In order to introduce the main ideas of the WM we consider the
second--order differential equation
\begin{equation}
L(y)=y^{\prime \prime }(x)+Q(x)y(x)=0  \label{eq:diffeq}
\end{equation}
If $y_{1}$ and $y_{2}$ are two solutions to this equation then we have
\begin{equation}
y_{1}L(y_{2})-y_{2}L(y_{1})=\frac{d}{dx}W(y_{1},y_{2})=0
\end{equation}
where
\begin{equation}
W(y_{1},y_{2})=y_{1}y_{2}^{\prime }-y_{2}y_{1}^{\prime }
\label{eq:Wronskian}
\end{equation}
is the Wronskian (or Wronskian determinant). If $y_{1}$ and $y_{2}$ are
linearly independent then the Wronskian (\ref{eq:Wronskian}) is a nonzero
constant. For practical purposes it is convenient to choose two solutions $%
C(x)$ and $S(x)$ that satisfy
\begin{equation}
C(x_{0})=S^{\prime }(x_{0})=1,\;C^{\prime }(x_{0})=S(x_{0})=0\;
\label{eq:C,S}
\end{equation}
at a given point $x_{0}$, so that $W(C,S)=1$ for all $x$.

In this paper we are interested in the dimensionless Schr\"{o}dinger
equation
\begin{equation}
-\frac{1}{2}\varphi ^{\prime \prime }(x)+v(x)\varphi (x)=\epsilon \varphi (x)
\label{eq:Schro_dim}
\end{equation}
where $\epsilon $ and $v(x)$ are the dimensionless energy and potential,
respectively. In earlier papers we have discussed a systematic way of
converting the standard Schr\"{o}dinger equation into its dimensionless form%
\cite{F11c,F11b,F11d} and we do not repeat it here. In order to facilitate
the discussion below we assume that $-\infty <x<\infty $, keeping in mind
that the main results can be extended to other cases if necessary. If $C(x)$
and $S(x)$ are two solutions satisfying (\ref{eq:C,S}) then we can write a
general solution as
\begin{equation}
\varphi (x)=A_{2}C(x)+B_{2}S(x)  \label{eq:phi_C_S}
\end{equation}
where, obviously,

\begin{eqnarray}
A_{2} &=&\varphi (x_{0})=W(\varphi ,S)\;  \nonumber \\
B_{2} &=&\varphi ^{\prime }(x_{0})=W(C,\varphi )  \label{eq:A2,B2,W}
\end{eqnarray}
For notation simplicity we write $C(x)$ and $S(x)$ instead of the more
detailed expressions $C(\epsilon ,x_{0},x)$ and $S(\epsilon ,x_{0},x)$,
respectively, that explicitly indicate the dependence of the linearly
independent solutions on the dimensionless energy $\epsilon $ and the chosen
coordinate point $x_{0}$.

It is well known that for an arbitrary value of the dimensionless energy $%
\epsilon $ the wavefunction behaves asymptotically as
\begin{equation}
\varphi (x)\rightarrow \left\{
\begin{array}{c}
A_{1}L_{c}(x)+B_{1}L_{d}(x),\,x\rightarrow -\infty \\
A_{3}R_{c}(x)+B_{3}R_{d}(x),\,x\rightarrow \infty
\end{array}
\right.  \label{eq:phi_asymp_gen}
\end{equation}
where $L$ and $R$ stand for left and right and $c$ and $d$ for convergent
and divergent, respectively. It means that, for arbitrary $\epsilon $, the
wavefunction is a linear combination of a convergent and a divergent
function when $|x|\rightarrow \infty $. If, for a particular value of $%
\epsilon $, $B_{1}=B_{3}=0$ then the resulting wavefunction is square
integrable. This condition determines the energies of the discrete spectrum.

It follows from the well known properties of the Wronskians\cite
{F11c,F11b,F11d} (and references therein) that
\begin{eqnarray}
B_{1}W(L_{c},L_{d})_{-} &=&A_{2}W(L_{c},C)_{-}+B_{2}W(L_{c},S)_{-}  \nonumber
\\
B_{3}W(R_{c},R_{d})_{+} &=&A_{2}W(R_{c},C)_{+}+B_{2}W(R_{c},S)_{+}
\end{eqnarray}
where the subscripts $-$ and $+$ indicate that the Wronskians are calculated
in the limits $x\rightarrow -\infty $ and $x\rightarrow \infty $,
respectively (where they become constants). Therefore, when $B_{1}=B_{3}=0$
we have a linear homogeneous system of two equations with two unknowns: $%
A_{2}$ and $B_{2}$. There will be nontrivial solutions provided that its
determinant vanishes
\begin{equation}
W(L_{c},C)_{-}W(R_{c},S)_{+}-W(R_{c},C)_{+}W(L_{c},S)_{-}=0
\label{eq:quant_cond_gen}
\end{equation}
The roots of this equation $\epsilon _{n}$, $n=0,1,\ldots $, are
the energies of the bound states.

When the potential is parity invariant
\begin{equation}
v(-x)=v(x)  \label{eq:v(x)_symm}
\end{equation}
and $x_{0}=0$ then $C(x)$ and $S(x)$ are even and odd functions,
respectively. In this case we have
\begin{eqnarray}
W(L_{c},S)_{-} &=&W(R_{c},S)_{+}  \nonumber \\
W(L_{c},C)_{-} &=&-W(R_{c},C)_{+}  \label{eq:W_symm}
\end{eqnarray}
and the determinant (\ref{eq:quant_cond_gen}) takes a simpler form: $%
W(R_{c},C)_{+}W(R_{c},S)_{+}=0$. We appreciate that the even and odd
solutions are clearly separate and their eigenvalues are given by
\begin{eqnarray}
W(R_{c},C)_{+} &=&0  \nonumber \\
W(R_{c},S)_{+} &=&0  \label{eq:quant_cond_even_odd}
\end{eqnarray}
respectively. Besides, we need to consider only the interval $0\leq x<\infty
$.

\section{The canonical--function method or Schr\"{o}dinger eigenvalue march}

\label{sec:canonical}

In order to compare the results of Sec.~\ref{eq:Wronskian} with
the SEM--CFM\cite{TL11,K82,TL10,TFL08} we simply note the
following equivalence between the main functions $\varphi
(x)\rightarrow y(x)$, $C(x)\rightarrow \alpha (x)$ and
$S(x)\rightarrow \beta (x)$. It follows from equation
(\ref{eq:A2,B2,W}) that
\begin{equation}
\frac{\varphi ^{\prime }(x_{0})}{\varphi (x_{0})}=\frac{W(C,\varphi )}{%
W(\varphi ,S)}=\frac{C\varphi ^{\prime }-C^{\prime }\varphi }{\varphi
S^{\prime }-\varphi ^{\prime }S}
\end{equation}
In the SEM--CFM one defines
\begin{eqnarray}
l_{-}(\epsilon ) &=&\lim_{x\rightarrow -\infty }\frac{W(C,\varphi )}{%
W(\varphi ,S)}=\frac{W(C,\varphi )_{-}}{W(\varphi ,S)_{-}}  \nonumber \\
l_{+}(\epsilon ) &=&\lim_{x\rightarrow \infty }\frac{W(C,\varphi )}{%
W(\varphi ,S)}=\frac{W(C,\varphi )_{+}}{W(\varphi ,S)_{+}}  \label{eq:l+l-}
\end{eqnarray}
and obtains the eigenvalues from the roots
of\cite{TL11,K82,TL10,TFL08}
\begin{equation}
F(\epsilon )=l_{+}(\epsilon )-l_{-}(\epsilon )=0  \label{eq:F(e)}
\end{equation}
Note that Eqs.~(\ref{eq:l+l-}) are identical to Eqs~(4) of Tannous and
Langlois\cite{TL11}. Therefore, the WM is quite similar to SEM except for
some slight differences that we will discuss later on.

For concreteness we assume that the acceptable solutions to the
dimensionless Schr\"{o}dinger equation (\ref{eq:Schro_dim}) satisfy the
boundary conditions
\begin{equation}
\lim_{|x|\rightarrow \infty }\varphi (x)=0
\end{equation}
In such a case it is customary to simplify the SEM--CFM equations
(\ref {eq:l+l-}) as follows\cite{TL11,K82,TL10,TFL08}
\begin{eqnarray}
l_{-}(\epsilon ) &=&\lim_{x\rightarrow -\infty }\frac{C(x)}{S(x)}  \nonumber
\\
l_{+}(\epsilon ) &=&\lim_{x\rightarrow \infty }\frac{C(x)}{S(x)}
\label{eq:l+l-_2}
\end{eqnarray}

In order to provide a rigorous proof for these equations we take
into account that
\begin{eqnarray}
C(x) &=&a_{1}L_{c}(x)+b_{1}L_{d}(x),\;x\rightarrow -\infty   \nonumber \\
S(x) &=&a_{1}^{\prime }L_{c}(x)+b_{1}^{\prime }L_{d}(x),\;x\rightarrow
-\infty   \nonumber \\
C(x) &=&a_{3}R_{c}(x)+b_{3}R_{d}(x),\;x\rightarrow \infty   \nonumber \\
S(x) &=&a_{3}^{\prime }R_{c}(x)+b_{3}^{\prime }R_{d}(x),\;x\rightarrow
\infty   \label{eq:C,S_asympt}
\end{eqnarray}
so that
\begin{eqnarray}
l_{-} &=&\lim_{x\rightarrow -\infty }\frac{C(x)}{S(x)}=\frac{b_{1}}{%
b_{1}^{\prime }}  \nonumber \\
l_{+} &=&\lim_{x\rightarrow \infty }\frac{C(x)}{S(x)}=\frac{b_{3}}{%
b_{3}^{\prime }}  \label{eq:l+l-_2b}
\end{eqnarray}
We can thus give a precise meaning to the word ``saturation''
often used in connection with the SEM--CFM\cite{TL11,TL10,TFL08}.
It simply points to the obvious fact that the divergent functions
dominate when $|x|$ is sufficiently large and $l_{\pm }$ tend to
the ratios of their coefficients in the expansion of the functions
$C(x)$ and $S(x)$. Besides, the WM gives us expressions for the
coefficients in equation (\ref{eq:C,S_asympt}) and their ratios
read
\begin{eqnarray}
\frac{b_{1}}{b_{1}^{\prime }} &=&\frac{W(C,L_{c})_{-}}{W(S,L_{c})_{-}}
\nonumber \\
\frac{b_{3}}{b_{3}^{\prime }} &=&\frac{W(C,R_{c})_{+}}{W(S,R_{c})_{+}}
\label{eq:ratios_b/b'}
\end{eqnarray}
Consequently equation (\ref{eq:F(e)}) leads to
(\ref{eq:quant_cond_gen}) that has been rigorously proved in
Sec.~\ref{sec:Wronskian}. Obviously, the right--hand sides of
equation (\ref{eq:l+l-_2b}) do not change if we substitute
$C^{\prime }(x)/S^{\prime }(x)$ for $C(x)/S(x)$ as pointed out by
Kobeissi\cite{K82} without giving a rigorous proof. More
precisely, if we substitute any pair of linearly independent
solutions for $C(x)$ and $S(x)$ in equations (\ref{eq:l+l-_2b})
and (\ref{eq:ratios_b/b'}) we should obtain a similar result. It
does not mean that all the pairs of solutions will be equally
efficient from a numerical point of view. We simply want to point
out that the WM provides a rigorous proof for the main equations
commonly used in the SEM--CFM.

If we require that $\varphi (x_{L})=\varphi (x_{R})=0$ for $x_{L}\ll
x_{0}\ll x_{R}$ then we have an homogeneous system of two equations with two
unknowns
\begin{eqnarray}
A_{2}C(x_{L})+B_{2}S(x_{L}) &=&0 \\
A_{2}C(x_{R})+B_{2}S(x_{R}) &=&0  \nonumber
\end{eqnarray}
that has nontrivial solutions only if
\begin{equation}
C(x_{L})S(x_{R})-C(x_{R})S(x_{L})=0  \label{eq:Dirichlet_left_right_det}
\end{equation}
This determinantal equation, derived earlier by Leubner\cite{L83}, is
equivalent to $l_{+}(\epsilon )-l_{-}(\epsilon )=0$ since $l_{+}(\epsilon
)=C(x_{R})/S(x_{R})$ and $l_{-}(\epsilon )=C(x_{L})/S(x_{L})$ in the right
and left asymptotic regions, respectively.

In the case of a symmetric potential and $x_{0}=0$ we have $C(-x)=C(x)$ and $%
S(-x)=-S(x)$ and the approximate eigenvalues are roots of the simpler
equation $C(x_{R})S(x_{R})=0$; that is to say $C(x_{R})=0$ or $S(x_{R})=0$
for the even or odd solutions, respectively.

\section{Examples}

\label{sec:examples}

As a first illustrative example, Tannous and Langlois\cite{TL11} considered
a particle of mass $m$ in a box with impenetrable walls at $x=0$ and $x=a$%
\begin{equation}
-\frac{\hbar ^{2}}{2m}\psi ^{\prime \prime }(X)=E\psi (X),\;\psi (0)=\psi
(a)=0  \label{eq:Schr_PB}
\end{equation}
as a model for an electron in a one--dimensional metallic rod of
finite length. The electron moves freely inside ($V(x)=0$) and
cannot escape from the rod ($V(x)=\infty $ if $x<0$ or $x>a$).
They applied the SEM and obtained the well known energies.
However, their choice of the model parameter $a$ is rather
atypical. Here, on the other hand, we transform the
Schr\"{o}dinger equation (\ref{eq:Schr_PB}) into a dimensionless
eigenvalue equation by means of the change of variables $X=ax$ and
$\varphi (x)=\sqrt{a}\psi (ax)$ that leads to $\varphi ^{\prime
\prime }(x)=-2\epsilon \varphi (x) $, where $\epsilon
=ma^{2}E/\hbar ^{2}$. Note that the boundary conditions for the
dimensionless solutions are $\varphi (0)=\varphi (1)=0$.

Tannous and Langlois mention the problem of defining proper
self--adjoint extensions of the operators for the infinite--well
potential\cite{BFV01}. This mathematical subtlety is of great
importance in the discussion of physical observables but it is not
an issue with regard to the calculation of the eigenvalues by
means of a numerical method like the SEM--CFM.

The particle in an infinite square well is suitable for the pedagogical
analysis of the performance of the shooting methods. If we consider a set of
discrete coordinate points $x_{j}=jh$, $j=0,1,\ldots ,N$ such that $%
x_{N}=Nh=1$ and define the approximate finite--difference first and second
derivatives
\begin{eqnarray}
\delta _{h}\varphi (x) &=&\frac{\varphi (x+h)-\varphi (x-h)}{2h}=\varphi
^{\prime }(x)+\frac{h^{2}\varphi ^{\prime \prime \prime }(x)}{6}+\ldots
\nonumber \\
\delta _{h}^{2}\varphi (x) &=&\frac{\varphi (x+2h)-2\varphi (x)+\varphi
(x-2h)}{4h^{2}}=\varphi ^{\prime \prime }(x)+\frac{h^{2}\varphi ^{IV}(x)}{3}%
+\ldots
\end{eqnarray}
then the Schr\"{o}dinger equation becomes a three--term recurrence
relation
\begin{equation}
\varphi _{j+2}+\left( 8h^{2}\epsilon -2\right) \varphi _{j}+\varphi _{j-2}=0
\end{equation}
where $\varphi _{j}=\varphi (x_{j})$ and $j=1,2,\ldots ,N-1$. On
substituting the solution $\varphi _{j}=e^{ij\theta }$ we obtain
\begin{equation}
\epsilon =\frac{1-\cos (2\theta )}{4h^{2}}
\end{equation}
Since $\varphi _{j}=e^{-ij\theta }$ is also a solution, then the general one
will be $\varphi _{j}=Ae^{ij\theta }+Be^{-ij\theta }$. From the boundary
conditions $\varphi _{0}=\varphi _{N}=0$ we obtain $\varphi _{j}=2Ai\sin
(j\theta )$, where $\theta =n\pi /N$, $n=1,2,\ldots ,N-1$ (note that $%
N\rightarrow \infty $ as $h\rightarrow 0$). Therefore, the approximate
eigenvalues are
\begin{equation}
\epsilon _{n}=\frac{1-\cos (2n\pi h)}{4h^{2}}=\frac{n^{2}\pi ^{2}}{2}-\frac{%
n^{4}\pi ^{4}h^{2}}{6}+\ldots
\end{equation}
This expression shows that the error in the numerical calculation of the
eigenvalues decreases quadratically with the step size $h$ and increases
with the quantum number $n$. A method is called $k^{th}$ order if its error
term is of order $h^{k+1}$\cite{PFTV86}. The naive shooting method just
outlined is first order. There are other well--known numerical integration
algorithms of greater order like the fourth--order Runge--Kuta method\cite
{PFTV86} that we will use in the calculations below.

\begin{figure}[H]
\begin{center}
\bigskip\bigskip\bigskip \includegraphics[width=9cm]{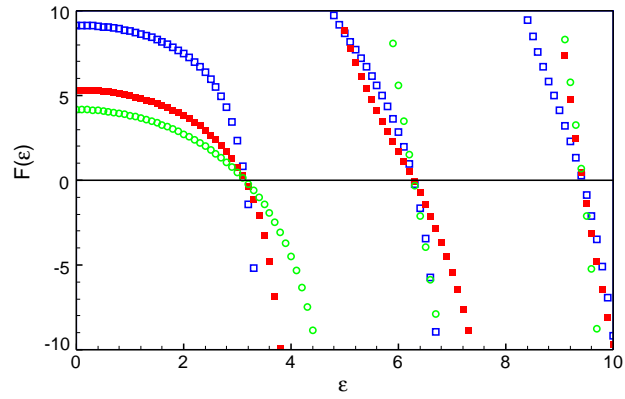}
\end{center}
\caption{Characteristic function $F(\epsilon)$ for the particle in a box for
$x_0 =1/8$ (blue squares), $x_0 =1/4$, (red solid squares) and $x_0 =2/5$
(green circles)}
\label{Fig:PB}
\end{figure}

Tannous and Langlois\cite{TL11} do not indicate the value of
$x_{0}$ chosen for their calculation on the infinite square well.
More precisely, they appear to be rather inconsistent about this
important point. They first state that if the potential is
symmetric in the interval $[x_{1},x_{2}]$ they choose
$x_{0}=(x_{1}+x_{2})/2$. However, in the discussion of the
boundary conditions, they say that if the potential is symmetric
in an interval of length $a$ they set $x_{1}=0$ and $x_{2}=a/2$ in
which case we expect a different value of this coordinate point:
$0<x_{0}<a/2$. Therefore, in order to illustrate the application
of the method to this trivial model we choose an arbitrary value
$0<x_{0}<1$ and do not exploit the symmetry of the equation about
$x=1/2$. The two linearly independent solutions that satisfy
Eq.~(\ref{eq:C,S}) are $C(x)=\cos [k(x-x_{0})]$ and
$S(x)=k^{-1}\sin [k(x-x_{0})]$, where $k=\sqrt{2\epsilon }$.
Obviously, in this case we do not have to bother about reaching
constant values of $l_{\pm }$ (saturation) because the coordinate
interval is finite. Upon substituting the Dirichlet boundary
conditions into equations (\ref{eq:F(e)}) and (\ref{eq:l+l-_2})
(adapted to the finite interval) we obtain
\begin{equation}
F(\epsilon )=l_{+}(\epsilon )-l_{-}(\epsilon )=-k\frac{\sin (k)}{\sin
(kx_{0})\sin [k(1-x_{0})]}  \label{eq:F(E)_PB}
\end{equation}
Fig.~\ref{Fig:PB} shows $F(\epsilon )$ for $0<\epsilon <10$ and $%
x_{0}=1/8,1/4,2/5$. Note that $F(\epsilon )$ vanishes at $k=n\pi $, $%
n=1,2,\ldots $ thus giving the well known eigenvalues disregarding the
chosen value of $x_{0}$. We also appreciate the effect of $x_{0}$ on the
form of the characteristic function $F(\epsilon )$.

Commonly, it is not difficult to derive approximate expressions
for the convergent and divergent asymptotic forms of the
wavefunction because they are straightforwardly determined by the
asymptotic behaviour of the potential $v(x)$. Therefore, it only
remains to have sufficiently accurate expressions for $C(x)$ and
$S(x)$ and their derivatives in order to obtain the eigenvalues by
means of the equations developed in sections \ref {sec:Wronskian}
and \ref{sec:canonical}. This problem is easily solved by means
of, for example, a suitable numerical integration
method\cite{PFTV86}. If $y(x)$ stands for either $C(x)$ or $S(x)$
then such an approach gives us its values at a set of points
$x_{0}-N_{L}h,\,x_{0}-N_{L}h+h,\ldots ,\,x_{0},\,x_{0}+h,\ldots
,\,x_{0}+N_{R}h$ where $N_{L}$ and $N_{R}$ are the number of steps
of size $h$ to the left and right of $x_{0}$, respectively. The
number of steps should be sufficiently large so that $y(x)$
reaches its asymptotic value at both $x_{L}=x_{0}-N_{L}h$ and
$x_{R}=x_{0}+N_{R}h$ and $h $ should be sufficiently small to
provide a good representation of $y(x)$.
The numerical integration methods also yield the derivative of the function $%
y^{\prime }(x)$ at the same set of points which facilitates the calculation
of the Wronskians. If the potential is parity invariant we only need to
integrate the Schr\"{o}dinger equation from $x_{0}=0$ to $x_{R}=N_{R}h$.

In Sec.~\ref{sec:canonical} we provided a rigorous basis for the
SEM--CFM that is one of the goals of this paper. In what follows
we illustrate those mathematical results by means of another
exactly solvable problem. For concreteness we choose

\begin{equation}
v(x)=-\frac{v_{0}}{\cosh ^{2}(x)},\;  \label{eq:v(x)_cosh}
\end{equation}
where $-\infty <x<\infty $. The allowed dimensionless energies are given by%
\cite{F11c,F99}
\begin{eqnarray}
\epsilon _{n} &=&-\frac{1}{2}(\lambda -1-n)^{2},\;n=0,1,\ldots \leq \lambda
-1  \nonumber \\
\lambda  &=&\frac{1}{2}\left( 1+\sqrt{1+8v_{0}}\right)
\label{eq:energ_exact}
\end{eqnarray}
and the spectrum is continuous for all $\epsilon >0$. It is clear that $%
\lambda \rightarrow 1$ as $v_{0}\rightarrow 0$ and there is only one bound
state when $1<\lambda <2$ ($0<v_{0}<1$). As $v_{0}$ increases more bound
states appear. As a result there are critical values of the potential
parameter for which $\epsilon _{n}=0$ that are given by the condition $%
\lambda _{n}=n+1$ or $v_{0,n}=\lambda _{n}(\lambda _{n}-1)/2=n(n+1)/2$.

Since $\lim_{|x|\rightarrow \infty }v(x)=0$ we have $R_{c}(x)\rightarrow
e^{-kx}$ and $R_{d}(x)\rightarrow e^{kx}$, where $k^{2}=-2\epsilon $ (we
only consider the interval $0\leq x<\infty $ because the potential is parity
invariant). Consequently, the allowed energies are determined by the
conditions
\begin{eqnarray}
W(R_{c},C)_{+} &=&\lim_{x\rightarrow \infty }\left[ C^{\prime
}(x)+kC(x)\right] e^{-kx}=0\;  \nonumber \\
W(R_{c},S)_{+} &=&\lim_{x\rightarrow \infty }\left[ S^{\prime
}(x)+kS(x)\right] e^{-kx}=0\;  \label{eq:quant_cond_even_odd_2}
\end{eqnarray}
for even and odd states, respectively.

Since the potential (\ref{eq:v(x)_cosh}) is parity invariant we integrate
the Schr\"{o}dinger equation from $x_{0}=0$ to $x_{R}=N_{R}h$. Fig.~\ref
{Fig:Wwell1} shows the Wronskians $W(R_{c},C)$ and $W(R_{c},S)$ and the
ratios $C(x)/S(x)$ and $W(R_{c},C)/W(R_{c},S)$ for the arbitrary values $%
\epsilon =-1$ and $v_{0}=2.5$. We appreciate that the ratios
$C(x)/S(x)$ and $W(R_{c},C)/W(R_{c},S)$ approach the same constant
value as $x\rightarrow \infty $ as proved in
Sec.~\ref{sec:canonical}. Note that the latter reaches the limit
at smaller coordinate values because the Wronskians take into
account the asymptotic form of the solution explicitly. In other
words, the WM requires less integration steps for the same
accuracy. We also appreciate that $x_{R}=5$ is large enough for
the WM and SEM--CFM calculations in this case. In order to compare
both approaches we find it reasonable to set $h=0.01$ and
$N_{R}=500$ in the fourth--order Runge--Kutta method\cite{PFTV86}
built in the computer algebra system Derive
(http://www.chartwellyorke. com/derive.html) that we use in all
our calculations. The numerical integration routine for the WM and
SEM--CFM is identical and the only difference is given by the
functions that we choose for the construction of $F(\epsilon )$.

Fig.~2 shows the Wronskians (\ref{eq:quant_cond_even_odd_2}) and
the ratios $C(x_{R})/S(x_{R})$ and $S(x_{R})/C(x_{R})$ for
$-10<\epsilon <0$ and $v_{0}=10$. We see that both approaches
yield the exact eigenvalues marked by black squares in the same
figure. The main difference is that the Wronskians change much
more smoothly than the canonical functions that cut the abscissae
axis sharply at the eigenvalues.

In many cases it is not difficult to derive the asymptotic form of
the convergent and divergent contributions to the wavefunction. As
another example consider the anharmonic oscillator
$v(x)=v_{2}x^{2}+v_{4}x^{4}$ ($v_4 >0$). A particular case is
given by the double well discussed by Tannous and
Langlois\cite{TL11}. If we introduce the ansatz $\varphi
(x)=e^{-f(x)}$ into the Schr\"{o}dinger equation and keep only the
dominant term we conclude that $R_{c}(x)\rightarrow \exp \left(
-\frac{\sqrt{2v_{4}}}{3}x^{3}\right) $ and $R_{d}(x)\rightarrow
\exp \left( \frac{\sqrt{2v_{4}}}{3}x^{3}\right) $ as $x\rightarrow
\infty $. On inserting these asymptotic expansions into the WM or
the SEM--CFM equations the saturation should appear at smaller
values of the coordinate as illustrated above by means of the
exactly solvable model potential (\ref{eq:v(x)_cosh}). Another
interesting example is provided by the radial part of the
dimensionless Schr\"{o}dinger equation for a central--field
potential $v(r)$:
\begin{equation}
-\frac{1}{2}\varphi ^{\prime \prime }(r)+\left[
\frac{l(l+1)}{2r^{2}}+v(r)\right] \varphi (r)=\epsilon \varphi (r)
\end{equation}
where $l=0,1,\ldots $ is the angular--momentum quantum number and
$\varphi (0)=0$. If we assume that $\lim_{r\rightarrow
0}r^{2}v(r)=0$ and insert the asymptotic behaviour $\varphi
(r)\rightarrow r^{s}$ at origin we conclude that
$L_{c}(r)\rightarrow r^{l+1}$ and $L_{d}(r)\rightarrow r^{-l}$ are
the regular and irregular contributions to the wavefunction.
Taking into consideration these asymptotic behaviours in the WM or
SEM--CFM equations will provide an advantage during the
integration from $r_{0}$ towards origin. According to the
discussion above one expects that it will not be necessary to
integrate too close to the origin to achieve saturation. On the
other hand, the asymptotic behaviour for $r\rightarrow \infty $ is
determined by the form of $v(r)$ as discussed above.

\section{Conclusions}

\label{sec:conclusions}

\begin{figure}[H]
\begin{center}
\bigskip\bigskip\bigskip \includegraphics[width=9cm]{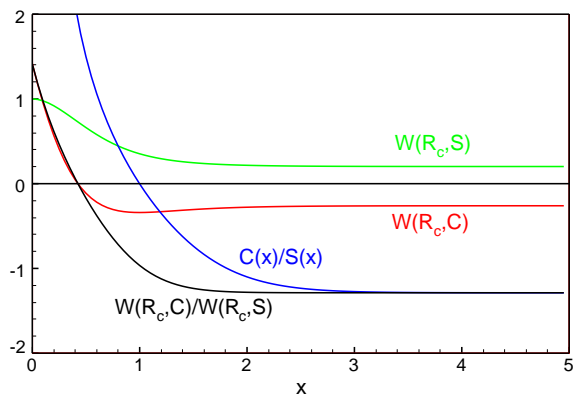}
\end{center}
\caption{Linearly independent solutions, Wronskians and their ratios for the
potential (\ref{eq:v(x)_cosh}) with $v_0=2.5$ and $\epsilon=-1$}
\label{Fig:Wwell1}
\end{figure}

The main purpose of this paper is to show that the Wronskians
provide a rigorous basis for the discussion of the SEM--CFM
equations as well as concepts like saturation. We have seen in
Fig.~\ref{Fig:Wwell1} that the Wronskians reach the asymptotic
value or saturation at smaller values of $|x|$ because they take
the asymptotic behaviour of the wavefunction explicitly into
account. However, the gain in numerical performance and efficiency
derived from this result is not as important as giving the
students the opportunity to discuss the asymptotic behaviour of
the wavefunction for a given quantum--mechanical problem. The most
general equations for both approaches developed in sections
\ref{sec:Wronskian} and \ref{sec:canonical} suggest that they are
essentially identical. If, for example, instead of substituting
the boundary conditions in the SEM--CFM function $F(\epsilon )$ we
substitute the correct asymptotic behaviour of the wavefunction
the SEM--CFM equation for the eigenvalues becomes the WM one. In
such a case that one finds it rather laborious to develop a
suitable analytical expression for the asymptotic behaviour of the
wave function at some singular point (as, for example, in the case
of the Lennard--Jones potential\cite{TFL08}) then it may be more
convenient to resort to the raw boundary conditions satisfied by
the wavefunction as discussed in Sec.~\ref{sec:canonical} (for
example, Eq. (\ref {eq:Dirichlet_left_right_det})). In other
words, we have the chance of using the SEM--CFM directly or
improving it by explicitly using the asymptotic behaviour of the
wavefunction.

In closing we want do discuss some points that may be of pedagogical value.
In the first place Tannous and Langlois\cite{TL11} state that the
characteristic equation $F(\epsilon )=0$ is not a matching condition but a
dispersion relation. The matching methods typically integrate the Schr\"{o}%
dinger equation inwards from left and right and require that $%
(y_{0})_{-}=(y_{0})_{+}$ and $(y_{0}^{\prime })_{-}=(y_{0}^{\prime })_{+}$
at an intermediate point $x_{0}$. Obviously, satisfying these equations is
equivalent to obtaining the roots of $F(\epsilon )=l_{+}(\epsilon
)-l_{-}(\epsilon )$. One can also integrate the logarithmic derivative $%
y^{\prime }(x)/y(x)$ inwards and obtain exactly the same equation.
On the other hand, the SEM--CFM integrates the canonical functions
(or normalized solutions) left and right from a given point
$x_{0}$ and then match the logarithmic derivative for each case at
that intermediate point. The difference between both strategies is
merely the direction of the integration (inwards or outwards). In
fact, Kobeissi\cite{K82} explicitly refers to the continuity
condition of the eigenfunction. The WM also proceeds outwards but
does not focus on the logarithmic derivative. Instead it makes use
of the Wronskians that become constant as the wavefunction
approaches the asymptotic region to derive equations for the
coefficients of the two linearly independent solutions.

The authors mention that the SEM evaluates the eigenvalues
directly and avoids losing accuracy associated with rapidly
oscillating wavefunctions of highly excited states. One should not
forget that the calculation of the canonical functions is
equivalent to the calculation of the wavefunction.
Kobeissi\cite{K82} explicitly shows the oscillatory behaviour of
such functions for an excited state of the Morse oscillator. It is
customary to state that the CFM does not calculate the
eigenfunctions explicitly\cite {TL11,K82,TL10,TFL08,K86} when it
is obvious that the approach already does it through the numerical
integration of the canonical functions. In other words, the
numerical integration of the Schr\"{o}dinger equation explicitly
calculates two oscillatory functions but their ratios (or the
appropriate Wronskians) do not reflect such oscillations as shown
in Fig.~\ref {Fig:Wwell1}.

In closing we want to discuss the statement that the SEM enables
`full determination of the spectrum in a single run'\cite{TL11}.
It is not clear what they mean because they calculate each root of
$F(\epsilon )=0$ by means of the secant method. This algorithm
requires the numerical evaluation of $F(\epsilon )$ several times
till $\epsilon $ is sufficiently close to the chosen root and each
calculation of $F(\epsilon )$ requires an outward integration of
the normalized solutions from $x_{0}$. In other words, just one
eigenvalue requires many integrations.

\begin{figure}[H]
\begin{center}
\bigskip\bigskip\bigskip \includegraphics[width=9cm]{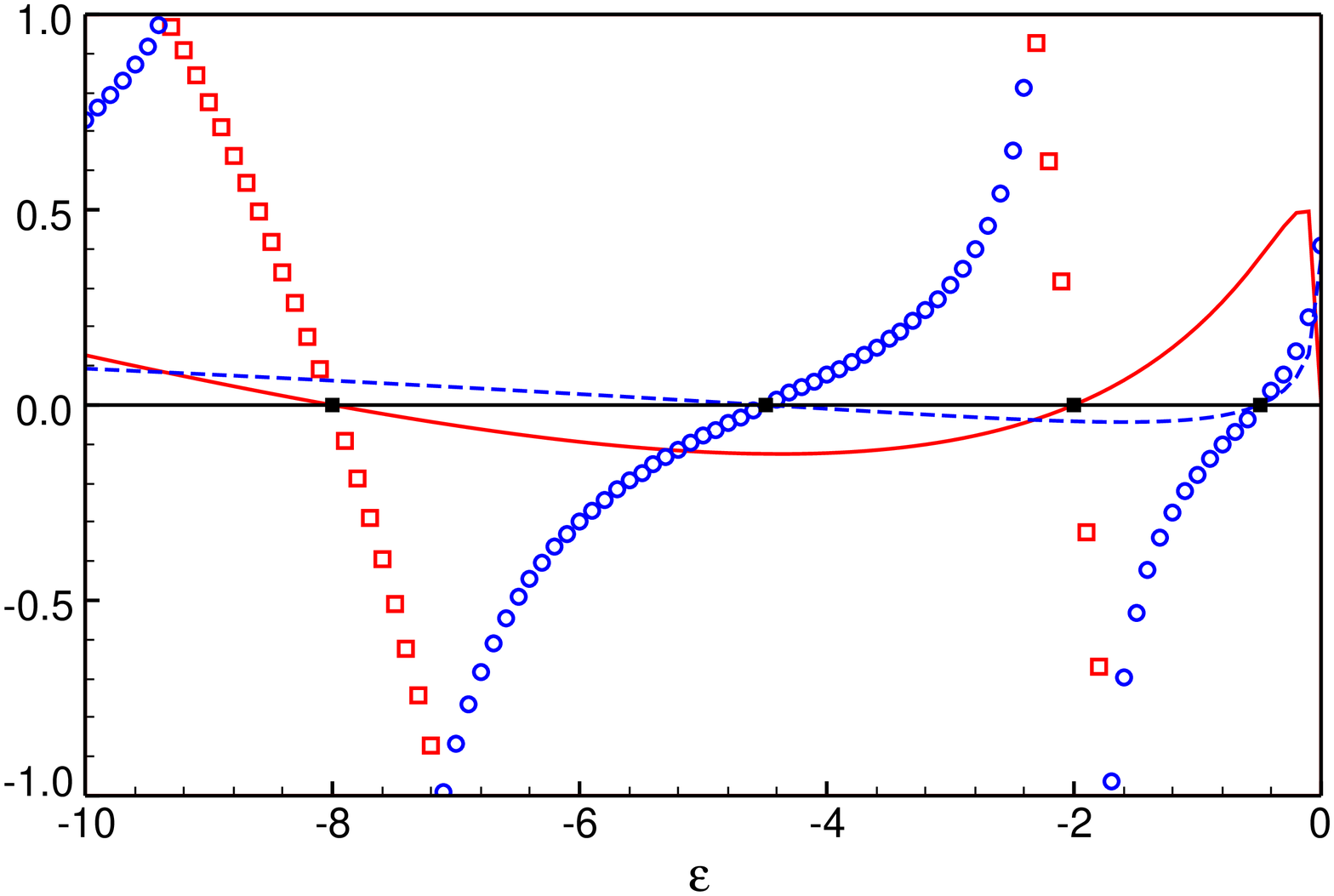}
\end{center}
\caption{$W(R_c,C)(x_R)$ (red solid line), $W(R_c,S)(x_R)$ (blue dashed
line), $C(x_R)/S(x_R)$ (red squares) and $S(x_R)/C(x_R)$ (blue circles) for
the potential (\ref{eq:v(x)_cosh}) with $v_0=10$}
\label{Fig:Wwell2}
\end{figure}


\begin{thebibliography}{99}
\bibitem{TL11}  Tannous C and Langlois J 2011 \textit{Eur. J. Phys.} \textbf{32} 1519.

\bibitem{K82}  Kobeissi H 1982 \textit{J. Phys. B} \textbf{693} 693.

\bibitem{TL10}Tannous C and Langlois J, Marching toward the eigenvalues: The
Canonical Function Method and the Schr\"odinger equation,
arXiv:1003.0184v1 [quant-ph].

\bibitem{TFL08}  Tannous C, Fakhreddine K, and Langlois J 2008 \textit{Phys.
Rep.} \textbf{467} 173.

\bibitem{L83}  Leubner C 1983 \textit{Int. J. Quantum Chem.} \textbf{24} 127.

\bibitem{K86}Kobeissi H 1986 {\it Int. J. Quantum Chem.} {\bf 29} 163.

\bibitem{T88}  Tellinghuisen J 1988 \textit{Int. J. Quantum Chem.} \textbf{34} 401.

\bibitem{F11c}  Fern\'{a}ndez F M 2011 \textit{Eur. J. Phys.} \textbf{32}
723 (arXiv:1101.3209v1 [quant-ph]).

\bibitem{F11b}  Fern\'{a}ndez F M 2011 \textit{Am. J. Phys.} \textbf{79}
877 (arXiv:1101.0957v1 [quant-ph]).

\bibitem{F11d}  Fern\'{a}ndez F M 2011 \textit{Eur. J. Phys.} (in press)
(arXiv:1107.4092v2 [quant-ph]).

\bibitem{BFV01}  Bonneau G, Faraut J, and Valent G 2001 \textit{Am. J. Phys.}
\textbf{69} 322.

\bibitem{PFTV86}  Press W H, Flannery B P, Teulosky S A, and Vetterling W T
1986 \textit{Numerical recipes. The art of scientific computing} (Cambridge
University Press, Cambridge).

\bibitem{F99}  Fl\"{u}gge S 1999 \textit{Practical Quantum Mechanics}
(Springer-Verlag, Berlin).
\end{thebibliography}
\end{document}